\documentclass{appolb}
\usepackage{epsfig}

\begin{document}
\title{Decays of Supersymmetric Particles: the program SUSY-HIT  
(SUspect-SdecaY-Hdecay-InTerface)}
\author{A.~Djouadi$^1$, M.M.~M\"uhlleitner$^2$\footnote{On leave from LAPTH, F-74941 Annecy-Le-Vieux Cedex, France.} and M.~Spira$^3$
\address{
$^1$ LPT, Universit\'e Paris-Sud, F-91405 Orsay, France, \\
$^2$ Physics Department, Theory Unit, CERN, CH-1211 Geneva 23, Switzerland, \\
$^3$ Paul Scherrer Institute, 5232 Villigen PSI, Switzerland } }

\maketitle

\vspace*{-7cm}
\hfill CERN-PH-TH/2006-200
\vspace*{6.8cm}
\vspace*{-6mm}
\begin{abstract}
We present the program package {\tt SUSY-HIT} for the computation of 
supersymmetric particle decays within the framework of the Minimal 
Supersymmetric extension of the Standard Model. The code is  based on two
existing programs {\tt HDECAY} and {\tt SDECAY} for the calculation of the 
decay widths and branching ratios of, respectively, the MSSM Higgs bosons and 
the SUSY particles, and calls a program for the calculation of the SUSY  
particle spectrum such as {\tt SuSpect}. Including all important higher order 
effects, the package allows the consistent calculation of the MSSM particle 
spectrum and decays with the presently highest level of  precision. 
\end{abstract}
\PACS{12.15-y, 12.60-i}
  
\section{Introduction}
\noindent Low energy Supersymmetry (SUSY) is considered as the prime candidate
for  physics beyond the Standard Model (SM) of the electroweak and strong 
interactions of elementary particles. It  will be testable at the Large Hadron
Collider (LHC) \cite{lhc} and later on at the International Linear Collider 
(ILC) \cite{ilc}. Once SUSY particles are discovered, their characteristics 
have to be determined with high accuracy to reconstruct the corresponding 
fundamental Lagrangian. This will be possible at the LHC at the few percent 
level, and a precision better than one percent will be achieved at the ILC. 
Therefore, theory  should provide calculational tools which match the expected 
high precision of the experiments. The codes for the determination of the SUSY 
particle spectrum and their decays must take into account corrections at the 
highest possible level. And since beyond leading order the results depend on  
many ingredients such as  the choice of the renormalization scheme and the 
definition of the input  parameters, an inherently consistent package to cover 
the whole program of calculating the spectra and decays of the new particles
at higher order is highly wishful.\\[0.2cm]
\noindent
In this context, we have developed the Fortran code {\tt SUSY-HIT} which
includes the codes {\tt HDECAY}  \cite{hdecay} and {\tt SDECAY}
\cite{sdecay} that allow the calculation of decay widths of both SUSY
particles and MSSM Higgs bosons, if the basic input parameters, the masses 
and the couplings of the particles are provided by a spectrum code\footnote{The
production of the SUSY and Higgs particles  of the MSSM  will be deferred to
other programs and ultimately, will be dealt with by  Monte--Carlo generators.
}. For the latter purpose, we have by  default used the program {\tt SuSpect}
\cite{suspect}, which can be replaced, however, by any spectrum  calculator
\cite{RGEcode} using the SUSY Les Houches Accord (SLHA) format \cite{slha} in 
the output file which will be read in by {\tt HDECAY} and {\tt SDECAY}.
\\[0.2cm]
In the following, we will first present the three programs separately: {\tt
SuSpect}, the spectrum calculator which is linked by default, and the decay
programs {\tt HDECAY} and {\tt SDECAY}, which constitute the heart of the
package. We recall how the codes work and which level of accuracy they 
contain. The program package {\tt SUSY-HIT} will be then briefly described.

\section{The spectrum calculator: the example of {\tt SuSpect}}
\noindent
\underline{Implementation of the MSSM:}
The Fortran code {\tt SuSpect} calculates the supersymmetric and Higgs particle
spectrum in the MSSM. It deals with the ``phenomenological MSSM'' with 22 free 
parameters defined either at a low or high energy scale, with the possibility 
of renormalization group evolution (RGE) to arbitrary scales, and with 
constrained models with universal boundary conditions at high scales. These 
are the minimal supergravity (mSUGRA), the anomaly mediated SUSY breaking 
(AMSB) and the gauge mediated SUSY breaking (GMSB) models. The basic 
assumptions of the most general possible MSSM scenario are {\it (a)} minimal 
gauge group, {\it (b)} minimal particle content, {\it (c)} minimal Yukawa 
interactions and R-parity conservation, {\it (d)} minimal set of soft SUSY 
breaking terms. Furthermore, {\it (i)} all soft SUSY breaking parameters are 
real (no CP-violation); {\it (ii)} the matrices for sfermion masses and 
trilinear couplings are diagonal; {\it (iii)} first and second sfermion 
generation universality is assumed. Here and in the following we refer the 
reader for more details to the user's manual \cite{suspect}. \\[0.2cm]
\underline{The general algorithm:} As for the calculation of the SUSY particle
spectrum in constrained MSSMs,  in addition to the choice of the input
parameters, the general algorithm  contains three main steps. These are {\it
(i)} the RGE of parameters back and  forth between the low energy scales, such
as $M_Z$ and the electroweak  symmetry breaking (EWSB) scale, and the
high-energy scale characteristic for  the various models; {\it (ii)} the
consistent implementation of (radiative)  EWSB; {\it (iii)} the calculation of
the pole masses of the Higgs bosons and  the SUSY particles, including the
mixing between the current eigenstates and  the radiative corrections when they
are important. The last step holds also  in unconstrained models. Here the
program mainly  follows the content and notations of Ref.~\cite{pbmz}, and for 
the leading two-loop  corrections to the Higgs boson masses the results given 
in Ref.~\cite{slavich} are used. The fulfillment of theoretical constraints as 
well as the agreement with high precision measurements can be also checked.
\\[0.2cm]
\underline{Necessary files and link to other programs:} The necessary files for
the use in {\tt SUSY-HIT} are the input file  {\tt suspect2.in}, the main
routine {\tt suspect2.f}, the routine  {\tt twoloophiggs.f}, which calculates
the Higgs masses, as well as  {\tt bsg.f} for the calculation of the $b\to
s\gamma$ branching ratio; other precision observables such as $(g-2)_\mu, 
\Delta\rho$ are calculated directly in {\tt suspect2.f}. Some of the files 
have been slightly modified  for the use in the package (for details see 
below). In the input file one can select the model to be investigated, the 
accuracy of the algorithm and the input data (SM particle masses and gauge 
couplings). At each run {\tt SuSpect} generates two output files: one easy to 
read, {\tt suspect2.out}, and the other in the SLHA format which can be read 
in by  {\tt HDECAY} and {\tt SDECAY}. 

\section{The Fortran code {\tt HDECAY}}
\noindent
\underline{Implemented decays:} The Fortran code {\tt HDECAY} calculates the 
decay widths and branching ratios of the SM Higgs boson, and of the neutral 
and charged Higgs particles of the MSSM according to the current theoretical 
knowledge \cite{know}. It includes all kinematically allowed decay channels 
with branching ratios larger than $10^{-4}$; in addition to the 2-body decays, 
also the loop-mediated, important multi-body and, in the MSSM, the cascade 
and SUSY decay channels are incorporated. More specifically, it includes: 
\vspace*{-0.3cm}

\begin{itemize}
\item[-] All relevant higher-order QCD corrections to decays into quarks
and to the quark loop mediated decays into gluons and photons \cite{qcdquark}. 
\vspace*{-0.2cm}

\item[-] Double off-shell decays of the CP-even Higgs bosons into massive 
gauge bosons, subsequently decaying into four massless fermions 
\cite{offgauge}.\vspace*{-0.2cm}

\item[-] All important below-threshold or 3-body decays: with off-shell heavy
top quarks; with one off-shell gauge boson as well as heavy neutral Higgs
decays with one off-shell Higgs boson \cite{higgs3body}. 
\vspace*{-0.2cm}

\item[-] In the MSSM the complete radiative corrections in the effective
potential approach with full mixing in the stop and sbottom sectors; it uses 
the RG improved values of the Higgs masses and couplings with  the relevant
next-to-leading-order corrections implemented \cite{effpot}. This will be 
needed to extract the Higgs self-couplings which the spectrum calculators 
do not provide by default.\vspace*{-0.2cm}

\item[-] In the MSSM, all the decays into SUSY particles when they are 
kinematically allowed \cite{susydecays} and all SUSY particles are included in 
the loop mediated $\gamma\gamma$ and $gg$ decay channels. In the gluonic 
and photonic decay modes the QCD corrections for quark \cite{quarkloops} and 
squark loops \cite{squarkloops} are implemented. 
\end{itemize}
\vspace*{-0.2cm}
\underline{Updates:}
{\tt HDECAY} has recently undergone a major upgrade. We have implemented the
SLHA format\footnote{M.~M\"uhlleitner has been added to the list of 
{\tt HDECAY} authors.}, so that the program can now read in any input 
file in this format and also provide the output for the Higgs decay widths and 
branching ratios in this accord. So, the program can now be easily linked to 
any spectrum or decay calculator. Two remarks are in order: \\[0.1cm]
\noindent
$i)$ {\tt HDECAY} calculates the higher order corrections to the Higgs boson 
decays in the $\overline{\mbox{MS}}$ scheme whereas all scale dependent 
parameters read in from an SLHA input file provided by a spectrum calculator 
are given in the $\overline{\mbox{DR}}$ scheme. Therefore, {\tt HDECAY} 
translates the input parameters from the SLHA file into the 
$\overline{\mbox{MS}}$ scheme where needed.\\[0.1cm]
\noindent
$ii)$  The SLHA parameter input file only includes the MSSM Higgs boson mass 
values, but not the Higgs self-interactions, which are needed in {\tt HDECAY}. 
For the time being, {\tt HDECAY} calculates the missing interactions internally
within the effective potential approach. This is not completely consistent 
with the values for the Higgs masses, since the spectrum  calculator does not
necessarily do it with the same method and level of  accuracy as {\tt HDECAY}.

\section{The Fortran code {\tt SDECAY}}
\noindent
\underline{Implemented decays:} 
The Fortran code {\tt SDECAY}, which has implemented the MSSM in the same way
as it is performed in {\tt SuSpect}, calculates the decay widths and branching 
ratios of all SUSY particles in the MSSM, including the most important higher 
order effects. More specifically: 
\vspace*{-0.3cm}

\begin{itemize}
\item[-]  In the default option, the usual 2-body decays for sfermions and 
gauginos are calculated at tree level. In GMSB models, the 2-body decays into  
the LSP  gravitino have been implemented.
\vspace*{-0.2cm}

\item[-] It provides the unique possibility of calculating the SUSY-QCD 
corrections to the decays involving coloured particles \cite{susyqcd}, which 
are very important in some cases. The bulk of the electroweak corrections is 
accounted for by using running parameters where  appropriate. \vspace*{-0.2cm}

\item[-] If the 2-body decays are kinematically closed, multibody decays will 
be dominant and {\tt SDECAY} calculates the 3-body decays of the gauginos, the 
gluino, the stops and sbottoms \cite{revisited}. \vspace*{-0.2cm}

\item[-] Loop-induced decays of the lightest top squark \cite{hikasa}, the 
next-to-lightest neutralino \cite{neutloop} and the gluino \cite{glloop} are 
included. \vspace*{-0.2cm}

\item[-] If the 3-body decays are kinematically forbidden, 4-body decays of 
the lightest stop \cite{4body} can compete with the loop-induced $\tilde{t}_1$ 
decay and have therefore been implemented. \vspace*{-0.2cm}

\item[-] Finally, the top decays within the MSSM have been included.
\vspace*{-0.2cm}
\end{itemize}
\underline{Updates:} {\tt SDECAY} has been recently updated and we take here 
the opportunity to announce the major changes. \\[0.1cm] 
\noindent
$i)$ For reasons of shortening the output file, only non-zero branching 
ratios are written out in the SLHA output file of the new version. \\[0.1cm]
\noindent
$ii)$ We have created common blocks for the branching ratios and total widths 
of the various SUSY particles. 

\section{The program package {\tt SUSY-HIT}}

\noindent
The following files are needed to run the program {\tt SUSY-HIT}:\\[0.1cm]
\underline{Spectrum files:} The spectrum can either be taken from any input 
file in the SLHA format or directly from {\tt SuSpect}. In the first case, 
{\tt SUSY-HIT} needs an SLHA input file which has to be named 
{\tt slhaspectrum.in}. In the latter case, {\tt SuSpect} provides this file
and we need all the necessary {\tt SuSpect} routines: {\tt suspect2.in}, 
{\tt suspect2.f}, {\tt twoloophiggs.f} and {\tt bsg.f}. \\[0.2cm]
\underline{Decay files:} For {\tt HDECAY} we need the main file 
{\tt hdecay.f} and for {\tt SDECAY} the main program file {\tt sdecay.f}. 
\\[0.2cm]
\underline{Input file:}
The flags which can be set in the original programs {\tt HDECAY} and {\tt
SDECAY} in their respective input file have been hard-coded for the use 
in {\tt SUSY-HIT}. {\tt SDECAY} will calculate by default the QCD 
corrections to the 2-body decays involving coloured particles, the multi-body
and the loop-induced decays, the top decays and the next-to-lightest SUSY 
particle decays in GMSB models. The running strong coupling constant and 
quark masses are calculated in the $\overline{\mbox{DR}}$ scheme at
the EWSB scale. In {\tt HDECAY}, higher order corrections and off-shell decays 
of all MSSM Higgs particles will be calculated, including the ones into SUSY 
particle final states. {\tt SUSY-HIT} thus needs only one input file called 
{\tt susyhit.in}. Here, the user can choose among two 
link options:\\[0.1cm]
$i)$ The three programs {\tt SuSpect}, {\tt HDECAY}, {\tt SDECAY} are
linked and, hence, {\tt SuSpect} provides the spectrum and SUSY breaking 
parameters at $M_{\rm SUSY}$.\\[0.1cm]
$ii)$ Only  {\tt HDECAY} and {\tt SDECAY} are linked, and the necessary inputs 
are taken from a file in the SLHA format provided by any spectrum calculator.
\\[0.1cm]
\noindent Furthermore, some parameters related to {\tt HDECAY} can be set, 
like some quark masses, the $W,Z$ total widths, some CKM matrix elements etc. 
All other necessary parameters are read in from {\tt slhaspectrum.in}.\\[0.2cm]
\noindent
\underline{Changes and how the package works:}
{\tt SuSpect}, {\tt HDECAY} and {\tt SDECAY} are lin\-ked via the SLHA format. 
Therefore, the name of the output file provided by {\tt SuSpect} has to be the 
same as the SLHA input file read in by {\tt HDECAY} and {\tt SDECAY}. We 
called it {\tt slhaspectrum.in}.  This is the major change made in the 
programs with respect to their original version. The complete list of changes 
can be found on the web page  given below, but we briefly comment on some of
them in the following.\\[0.2cm]
\noindent \underline{{\tt SDECAY}:}
It is the main program and now reads in {\tt susyhit.in} and calls 
{\tt HDECAY}. It passes the necessary parameters from {\tt susyhit.in} to 
{\tt HDECAY} via a newly created common block called {\tt SUSYHITIN}. As 
before, it calls {\tt SuSpect} in case the spectrum is taken from there. The 
SLHA parameter and spectrum input file {\tt slhaspectrum.in} is read in by 
both {\tt HDECAY} and {\tt SDECAY}. The output file created by {\tt SDECAY} at 
each run is called {\tt susyhit\_slha.out} if it is in the SLHA format or 
simply {\tt susyhit.out} if it is in an output format easy to read. \\[0.2cm]
\noindent \underline{{\tt SuSpect}:}
Some subroutines within the file {\tt twoloophiggs.f} which are also used in 
{\tt hdecay.f} have been renamed in order to avoid clashes. \\[0.2cm]
\noindent \underline{{\tt HDECAY}:}
It has become a subroutine which is called by {\tt SDECAY}. In order to keep
the package as small as possible, only one routine calculating the Higgs boson 
masses and Higgs self-couplings has been retained in {\tt HDECAY} to extract 
the Higgs self-interaction strengths not provided by the spectrum calculators. 
A new common block {\tt SUSYHITIN} has been created to take over necessary 
parameters from {\tt SDECAY}. {\tt HDECAY} does not create any output file 
within the package. \\[0.2cm]
\noindent
\underline{Remarks:} The $b$-quark mass given in the SLHA input file is the 
running mass $m_b(m_b)^{\overline{\rm MS}}$ in the 
$\overline{\mbox{MS}}$ 
scheme, which is the mass measured by the experiments. The three programs 
calculate internally the $b$-quark pole mass from this value. {\tt HDECAY} and 
{\tt SuSpect} do this differently. {\tt SDECAY} takes the same value for the 
pole mass as provided by {\tt SuSpect}. So slightly different $b$-quark pole 
masses will be used in the decay calculations. The difference in the pole mass 
is less than 3\%. The user should furthermore keep in mind that the Higgs 
masses taken from the spectrum calculator and the Higgs self-couplings which 
are used within {\tt HDECAY} may be slightly inconsistent, since these masses 
and couplings are related, but not provided together by the SLHA input file.
\\[0.2cm]
\noindent \underline{Web page:} We have created a web page at the following
url address:\\
\centerline{{\tt http://lappweb.in2p3.fr/$\sim$muehlleitner/SUSY-HIT/}}
There the user can download all files necessary for the program package
as well as a {\tt makefile} for compiling the programs. We use the newest 
versions of the various programs which will be updated regularly. Short 
instructions are given how to use the programs. A file with updates and 
changes is provided. Finally, some examples of output files are given.

\section{Summary}

\noindent We have briefly presented the program package {\tt SUSY-HIT} for the 
calculation of the particle spectrum in the MSSM and the decay widths and 
branching ratios of the SUSY particles and the Higgs bosons. The core of the 
package are the already available decay codes {\tt SDECAY} and {\tt HDECAY}. 
They are linked by default to the MSSM spectrum calculator {\tt SuSpect} which,
however, can be replaced by any  other spectrum code giving the output in the
SLHA format.  The program provides the decay widths and branching ratios of
the SUSY particles and  neutral/charged Higgs bosons of the MSSM, as well 
as the additional decays  of the top quark, at the presently  highest level of
precision.  Hence, except for the production processes which ultimately must be
dealt with by Monte--Carlo generators (the link with the latter can be easily
made through the SLHA), the program {\tt SUSY-HIT} provides all the necessary 
information needed for the search and the study of the new particles at 
high--energy colliders. \\[0.2cm]
\noindent  
The package is self-contained, easy to use, flexible and
rather fast. All files and further information can be found on the related
homepage. {\tt SUSY-HIT} will be upgraded regularly to keep up with the 
experimental needs and theoretical developments. Suggestions for improving 
and adapting the code from theorists and experimentalists are highly welcome.

\vspace*{0.4cm}
\noindent \underline{\bf Acknowledgements:}
We would like to thank the other authors of the original programs 
{\tt HDECAY}, {\tt SDECAY} and {\tt SuSpect}, J.~Kalinowski, J.-L.~Kneur, 
Y.~Mambrini and G.~Moultaka for their support in the past.



\begin{thebibliography}{99}
\bibitem{lhc}
CMS Technical Design Report, Report CERN/LHCC 2006-001; ATLAS 
Collaboration, Technical Design Report, Vols. 1 and 2, CERN-LHCC-99-14 and 
CERN-LHC-99-15. See also the proceedings of the les Houches workshops,
hep-ph/0002258, hep-ph/0203056 and hep-ph/0406152.
%
\bibitem{ilc}
E.~Accomando et al., Phys. Rep. {\bf 299} (1998) 1; J.A. Aguilar-Saavedra et
al., TESLA-TDR, hep-ph/0106315; for informations on the ILC, see the web site: 
{\tt http://www.interactions.org/linearcollider/}. 
%
\bibitem{hdecay} A.\,Djouadi, J.\,Kalinowski and M.\,Spira, Comput. Phys. 
Commun. {\bf 108} (1998) 56.
%
\bibitem{sdecay} M.~M\"uhlleitner, A.~Djouadi and Y.~Mambrini, Comput. Phys.
Commun. {\bf 168} (2005) 46; M.~M\"uhlleitner, Acta Phys. Polon. {\bf B35}
(2004) 2753.
%
\bibitem{suspect} A.~Djouadi, J.-L.~Kneur and G.~Moultaka, hep-ph/0211331.
%
\bibitem{RGEcode} Other publically available RGE codes are for 
instance: {\tt ISAJET}, H.~Baer, F.E.~Paige, S.D.~Protopopescu and X.~Tata, 
hep-ph/0001086; 
{\tt SOFTSUSY}, B.~Allanach, Comput. Phys. Commun. {\bf 143} (2002) 305; 
{\tt SPHENO}, W.~Porod, Comput.\ Phys.\ Commun. {\bf 153} (2003) 275.
%
\bibitem{slha} P.~Skands et al., JHEP {\bf 0407} (2004) 036.
%
\bibitem{pbmz} D.\,Pierce, J.\,Bagger, K.\,Matchev and R.\,Zhang, Nucl.
Phys. {\bf B491} (1997) 3.
%
\bibitem{slavich} G.~Degrassi, P.~Slavich and F.~Zwirner, Nucl. Phys. 
{\bf B611} (2001) 403; 
A.~Brignole et al., Nucl. Phys. {\bf B631} (2002) 195 and {\bf B643} (2002) 79;
A.~Dedes, G.~Degrassi and P.~Slavich Nucl. Phys. {\bf B672} (2003) 144; 
B. Allanach et al., JHEP {\bf 0409} (2004) 044.
%
\bibitem{know} A.~Djouadi, hep-ph/0503172 and hep-ph/0503173; 
M.~Gomez-Bock et al., J. Phys. Conf. Ser. {\bf 18} (2005) 74 [hep-ph/0509077].
%
\bibitem{qcdquark} See e.g. A.~Djouadi, M.~Spira and P.M.~Zerwas, 
Z. Phys. {\bf C70} (1996) 427.
%
\bibitem{offgauge} See e.g. R.N.~Cahn, Rep. Prog. Phys. {\bf 52} (1989) 389.
%
\bibitem{higgs3body} A.~Djouadi, J.~Kalinowski and P.M.~Zerwas, Z. Phys.
{\bf C70} (1996) 437; S.~Moretti and W.J.~Stirling, Phys. Lett. {\bf B347}
(1995) 291, (E) {\bf B366} (1996) 451.
%
\bibitem{effpot} M.\,Carena, M.\,Quiros and C.E.M.\,Wagner, Nucl. Phys. 
{\bf B461} (1996) 407.
%
\bibitem{susydecays} A.\,Djouadi, J.\,Kalinowski, P.\,Ohmann and P.\,Zerwas,
Z. Phys. {\bf C74} (1997) 93.
%
\bibitem{quarkloops}
M.\,Spira, A.\,Djouadi, D.\,Graudenz and P.\,Zerwas, Nucl. Phys. {\bf B453}
(1995) 17; T.\,Inami, T.\,Kubota and Y.\,Okada, Z. Phys. {\bf C18} (1983) 69;
A.\,Djouadi, M.\,Spira and P.\,Zerwas, Phys. Lett. {\bf B264} (1991) 440;
K.\,Chetyrkin, B.\,Kniehl and M.\,Steinhauser, Phys. Rev. Lett. {\bf 78}
(1997) 594.
%
\bibitem{squarkloops} S.~Dawson, A.~Djouadi and M.~Spira, Phys. Rev. Lett.
{\bf 77} (1996) 16.
%
\bibitem{susyqcd} S.~Kraml et al., Phys. Lett. {\bf B386} (1996) 175; 
A.~Djouadi, W.~Hollik and C.~J\"unger, Phys. Rev. {\bf D55} (1997) 6975;
A.~Arhrib et al., Phys. Rev. {\bf D57} (1998) 5860;
A.~Bartl et al. Phys. Rev. {\bf D59} (1999) 115007, Phys. Lett. {\bf B419}
(1998) 243 and Phys. Lett. {\bf B435} (1998) 118; W.~Beenakker, R~H\"opker and
P.M.~Zerwas, Phys. Lett. {\bf B378} (1996) 159; W.~Beenakker, R.~H\"opker, 
T.~Plehn and P.M.~Zerwas, Z. Phys. {\bf C75} (1997) 349.
%
\bibitem{revisited} A.~Djouadi, Y.~Mambrini and M.~M\"uhlleitner, Eur. Phys. J.
{\bf C20} (2001) 563 and references therein.
%
\bibitem{hikasa} K.I.~Hikasa and M.~Kobayashi, Phys. Rev. {\bf D36} (1987) 
724.
%
\bibitem{neutloop} H.E.~Haber and D.~Wyler, Nucl. Phys. {\bf B323} (1989) 
267; S.~Ambrosanio and B.~Mele, Phys. Rev. {\bf D53} (1996) 2541 and {\bf D55}
(1997) 1399; H.~Baer and T.~Krupovnickas, JHEP {\bf 0209} (2002) 038.
%
\bibitem{glloop} E.~Ma and G.G.~Wong, Mod. Phys. Lett. {\bf A3} (1988) 1561;
R.~Barbieri et al., Nucl. Phys. {\bf B301} (1988) 15; H.~Baer, X.~Tata and 
J.~Woodside, Phys. Rev. {\bf D42} (1990) 1568.
%
\bibitem{4body} C.~Boehm, A.~Djouadi and Y.~Mambrini, Phys. Rev. {\bf D61} 
(2000) 095006.
%
\end{thebibliography}
\end{document}